# Enhancement of Pure Spin Currents in Spin Pumping $Y_3Fe_5O_{12}$/Cu/metal Trilayers through Spin Conductance Matching

Chunhui Du†, Hailong Wang†, Fengyuan Yang*, and P. Chris Hammel*

Department of Physics, The Ohio State University, Columbus, OH, 43210, USA

†These authors made equal contributions to this work

*Emails: fyyang@physics.osu.edu, hammel@physics.osu.edu

Spin transport efficiency in heterostructures depends on the spin conductances of each constituent and their interfaces. We report a comparative study of spin pumping in $Y_3Fe_5O_{12}$/Cu/Pt and $Y_3Fe_5O_{12}$/Cu/W trilayers. Surprisingly, the insertion of a Cu interlayer between $Y_3Fe_5O_{12}$ and W substantially improves (over a factor of 4) the spin current injection into W while similar insertion between $Y_3Fe_5O_{12}$ and Pt degrades the spin current. This is a consequence of a much improved $Y_3Fe_5O_{12}$/Cu spin mixing conductance relative to that for $Y_3Fe_5O_{12}$/W. This result implies the possibility of engineered heterostructures with matching spin conductances to enable optimal spin transport efficiency.

Conventional electronic devices operate via flow of electrical charges. The decoupling of spin and charge currents in ferromagnetic resonance (FMR) driven spin pumping offer the potential to enable low energy cost, high efficiency spintronics with implications for both next generation computing [1, 2] and global energy consumption [3]. Spin pumping, driven thermally as well as by FMR, is being widely used to generate pure spin currents from ferromagnets (FM) into normal metals (NM) [4-18]. The efficiency of spin pumping is largely determined by the spin mixing conductance $g_{\uparrow\downarrow}$ [4, 5] of the FM/NM interface. Typically, the NM is chosen to be a spin-sink—Pt, W or Ta with large inverse spin Hall effect (ISHE), while lighter metals such as Cu are rarely used. Various FM/NM or FM/$NM_1$/$NM_2$ structures have been extensively studied by both FMR spin pumping [6-16] and spin Seebeck measurements [17, 18]. Due to its low magnetic damping and insulating nature, $Y_3Fe_5O_{12}$ (YIG) is very attractive for microwave applications and spin pumping [6, 7, 11-13, 15, 18-21]. It will be important to understand the spin transmissivity of multilayers in order to enable the use of intervening layers for optimizing spin transport properties. Cu interlayers were recently used to eliminate proximity induced ferromagnetism in YIG/Pt [18, 22-24]. A quantitative understanding of spin current generation and transport in YIG/Cu and YIG/Cu/NM heterostructures, especially the spin mixing or spin conductance at each interface, is still lacking.

We use epitaxial YIG(20 nm) films grown on $Gd_3Ga_5O_{12}$ (111) by off-axis sputtering; these enable mV-level ISHE voltages in YIG/Pt and YIG/W spin pumping measurements [11, 12, 25]. Figure 1a shows an x-ray diffraction (XRD) scan of a YIG film with clear Laue oscillations and a rocking curve full-width-at-half-maximum (FWHM) of 0.0073° (inset to Fig. 1a). Figures 1b-1d show atomic force microscopy (AFM) images of a 20-nm YIG film, a YIG/Cu(5 nm) and a YIG/Cu(20 nm) bilayer with a root-mean-square (rms) roughness of 0.15, 0.19 and 0.22 nm, respectively, indicating the smooth surfaces of the YIG films and the Cu layers.

Our FMR ($f$ = 9.65 GHz) spin pumping measurements are performed at room temperature on YIG(20 nm)/Cu($t_{Cu}$)/Pt(5 nm) and YIG(20 nm)/Cu($t_{Cu}$)/W(5 nm) trilayers where the Cu thickness $t_{Cu}$ is varied from 0 to 20 nm. All metal layers are deposited by off-axis, ultrahigh vacuum sputtering [11, 12, 25]. Samples ~1 mm wide and ~5 mm long are placed in the center of an FMR cavity in a DC magnetic field (***H***) applied in the *xz*-plane, as shown in the schematic in Fig. 2a. At resonance the precessing YIG magnetization transfers angular momentum to the conduction electrons in Cu. Since $t_{Cu}$ is much smaller than the spin diffusion length ($\lambda_{SD}$), spin accumulation in the Cu spacer drives a spin current $J_s$ into the Pt or W layer. As a consequence of the ISHE, the spin current $J_s$ into the Pt or W is converted into a charge current, resulting in a voltage ($V_{ISHE}$) along the *y*-axis. Figures 2b and 2c show $V_{ISHE}$ vs. $H$ spectra for YIG/Pt and YIG/W bilayers at $\theta_H$ = 90° and 270° (both in-plane fields) at rf input power $P_{rf}$ = 200 mW, which gives $V_{ISHE}$ = 552 μV and -2.04 mV, respectively. The sign reversal for YIG/W reflects the opposite spin Hall angles of W and Pt. Figures 2d and 2e show the $V_{ISHE}$ vs. $H$ spectra for YIG/Cu(20 nm)/Pt and YIG/Cu(20 nm)/W trilayers. The peak values decrease to $V_{ISHE}$ = 1.21 μV and -16.3 μV as compared to YIG/Pt and YIG/W with direct contact, respectively.

To compare the effect of the Cu spacer on $J_s$, the spin current detected in Pt or W in the two trilayer systems, we show in Figs. 2f and 2g the $t_{Cu}$ dependence of $J_s$ normalized by $J_s(0)$, the spin current detected in the YIG/Pt and YIG/W bilayers without Cu interlayer. Spin current $J_s$ can be calculated from [26, 27],

$$J_s = \frac{2e}{\hbar}\frac{V_{ISHE}}{\theta_{SH}\lambda_{SD}\tanh(\frac{t_{NM}}{2\lambda_{SD}})wR}, \quad (1)$$

where $t_{NM}$ and $\theta_{SH}$ are the thickness and spin Hall angle of Pt or W, $R$ and $w$ the total resistance and width of the trilayers, respectively. The total resistance for the YIG/Pt bilayer sample is 487.4 Ω. As the Cu interlayer is inserted between YIG and Pt with increasing thickness, the total resistance $R$ decreases and reaches 7.7 Ω at 20-nm Cu due to the shunting effect of Cu. For the YIG/W bilayer, the resistance is 6045.6 Ω, while it decreases to 9.3 Ω for the YIG/Cu(20 nm)/W trilayer. We note that the term $\theta_{\mathrm{SH}}\lambda_{\mathrm{SD}}\tanh(\frac{t_{\mathrm{NM}}}{2\lambda_{\mathrm{SD}}})$ in Eq. (1) is cancelled in the normalized spin current $J_{\mathrm{s}}/J_{\mathrm{s}}(0)$. Taking the variation of total resistance $R$ and width $w$ into account, we show in Figs. 2f and 2g that with the insertion of Cu, the normalized spin current $J_{\mathrm{s}}/J_{\mathrm{s}}(0)$ initially decreases dramatically at $t_{\mathrm{Cu}} \leq 5$ nm, then increases and eventually reaches a plateau at $t_{\mathrm{Cu}} \geq 10$ nm. However, YIG/Cu/Pt and YIG/Cu/W show opposite plateau values of $J_{\mathrm{s}}/J_{\mathrm{s}}(0)$: for YIG/Cu/Pt, $J_{\mathrm{s}}$ is only about 20-25% of $J_{\mathrm{s}}(0)$, while for YIG/Cu/W, $J_{\mathrm{s}}$ is 4.5 times of $J_{\mathrm{s}}(0)$. The initial decrease of $J_{\mathrm{s}}$ at $t_{\mathrm{Cu}} \leq 5$ nm may be related to the much higher resistivity of thin Cu layers due to finite size effect which could induce significant spin flipping [4, 5]. At $t_{\mathrm{Cu}} \geq 10$ nm, the resistivity of Cu layers decreases significantly and spin accumulation in Cu occurs. A fraction of the accumulated spins in Cu diffuse back into YIG while the remainder is transmitted into the Pt or W, producing an ISHE signal. The ratio between these two fractions is determined by the interfacial spin mixing conductances $g_{\mathrm{YIG/Cu}}^{\uparrow\downarrow}$ and spin conductance $g_{\mathrm{Cu/Pt}}$ and $g_{\mathrm{Cu/W}}$.

To uncover the mechanism behind the different spin pumping behavior of the two trilayers systems, we need to determine the spin mixing conductances from the enhancement of Gilbert damping $\alpha$ due to spin pumping. We obtain this by measuring the frequency dependencies of the FMR linewidth $\Delta H$ using a broadband microstrip transmission line. In all cases the linewidth increases linearly with frequency from 10 to 20 GHz (Fig. 3) following [28],

$$\Delta H = \Delta H_{\mathrm{inh}} + \frac{4\pi\alpha f}{\sqrt{3}\gamma}, \qquad (2)$$

where $\Delta H_{\mathrm{inh}}$ is the inhomogeneous broadening and $\gamma$ is the gyromagnetic ratio. We define the enhancement of Gilbert damping due to spin pumping, $\alpha_{\mathrm{sp}} = \alpha_{\mathrm{YIG/Cu/NM}} - \alpha_{\mathrm{YIG}}$ for trilayers and $\alpha_{\mathrm{sp}} = \alpha_{\mathrm{YIG/NM}} - \alpha_{\mathrm{YIG}}$ for bilayers.

Figure 3a and Table I show that $\alpha_{\mathrm{sp}}$ = (2.1 ± 0.1) × $10^{-3}$ for YIG/Pt approximately doubles that for YIG/Cu(20 nm)/Pt, (1.1 ± 0.1) × $10^{-3}$, in agreement with the previous study on Py/Cu/Pt [14] and YIG/Cu/Pt multilayers [23]. However, the order is reversed for YIG/Cu/W: $\alpha_{\mathrm{sp}}$ = (2.1 ± 0.2) × $10^{-3}$ for YIG/Cu(20 nm)/W is more than 3 times larger than the value of (6.3 ± 0.6) × $10^{-4}$ for YIG/W (Fig. 3b). Cavity FMR measurements also confirm this trend as shown in Figs. 3d and 3e. The linewidth change for YIG/Cu(20 nm)/Pt compared to the bare YIG is 6.0 Oe which is smaller than the value of 12.6 Oe for YIG/Pt. However, for YIG/Cu(20 nm)/W trilayer, the linewidth change is 13.1 Oe which is larger than the value of 6.4 Oe for YIG/W. This implies significant difference at the interfaces in YIG/Cu/Pt and YIG/Cu/W compared to YIG/Pt and YIG/W. In order to understand this behavior, we need to determine the spin mixing conductances of $g_{\mathrm{YIG/Cu}}^{\uparrow\downarrow}$, $g_{\mathrm{YIG/Pt}}^{\uparrow\downarrow}$ and $g_{\mathrm{YIG/W}}^{\uparrow\downarrow}$, and the spin conductances of $g_{\mathrm{Cu/Pt}}$ and $g_{\mathrm{Cu/W}}$.

Tserkovnyak *et al*. provide a theory for quantitative analysis of interfacial spin mixing conductance [4, 5] for metallic FM/NM bilayers or trilayers where the real part of $g_{\uparrow\downarrow}$ is dominant. For YIG/NM interfaces, it has been reported recently that the imaginary part of $g_{\uparrow\downarrow}$ is also negligibly small [29, 30]. Thus, this theory is applicable for YIG/NM systems as well. The spin pumping induced Gilbert damping enhancement can be expressed as [4, 5, 7, 31-33],

$$\alpha_{\mathrm{sp}} = \frac{g\mu_{\mathrm{B}}}{4\pi M_{\mathrm{s}} t_{\mathrm{YIG}}} g_{\mathrm{eff}}^{\uparrow\downarrow}, \quad (3)$$

where $g_{\mathrm{eff}}^{\uparrow\downarrow}$ is the effective spin mixing conductance at the YIG/NM interfaces which includes the spin current backflow driven by spin accumulation, $g$, $\mu_{\mathrm{B}}$, $M_{\mathrm{s}}$ ($4\pi M_{\mathrm{s}}$ = 1794 Oe [11]) and $t_{\mathrm{YIG}}$ are the Landé $g$ factor, Bohr magneton, saturation magnetization and thickness of the YIG layer, respectively. Rapid spin flips relax the injected spin very quickly in good spin sink materials such as Pt and W, preventing any backflow that results from spin accumulation, thus $g_{\rm eff}^{\uparrow\downarrow} \approx g^{\uparrow\downarrow}$. However, in NMs with slower relaxation, such as Cu [4, 5],

$$g_{\rm eff}^{\uparrow\downarrow} = g_{\rm YIG/NM}^{\uparrow\downarrow}[1 + g_{\rm YIG/NM}^{\uparrow\downarrow} \frac{1}{4\sqrt{\frac{\epsilon}{3}}\tanh(\frac{t_{\rm NM}}{\lambda_{\rm SD}})g_{\rm NM}^{\rm Sh}}]^{-1}, \quad (4)$$

where $g_{\rm YIG/NM}^{\uparrow\downarrow}$, $g_{\rm NM}^{\rm Sh}$, and $\epsilon$ are the "intrinsic" spin mixing conductance of the YIG/NM interface, Sharvin conductance of the NM, and the spin-flip probability of the NM [4, 5].

The spin accumulation in Cu leads to backflow into the YIG so $g_{\rm eff}^{\uparrow\downarrow}$ is much smaller than $g_{\rm YIG/Cu}^{\uparrow\downarrow}$ depending on $t_{\rm NM}$ and $\epsilon$. For YIG/Cu/NM trilayers where the NM is a spin sink, in the limit of vanishing spin flip in Cu [31, 32], the effective spin mixing conductance $g_{\rm eff,trilayer}^{\uparrow\downarrow}$ of the trilayer is simply the serial contributions of the two interfaces and the spin resistance of Cu,

$$\frac{1}{g_{\rm eff,trilayer}^{\uparrow\downarrow}} = \frac{1}{g_{\rm YIG/Cu}^{\uparrow\downarrow}} + R_{\rm Cu} + \frac{1}{g_{\rm Cu/NM}}, \quad (5)$$

Where the spin resistance $R_{\rm Cu} = \frac{2e^2 t_{\rm Cu}}{h\sigma}$ ($\sigma$ is the electrical conductivity) [4, 5]. Using measured resistivity of $4.0 \times 10^{-8}$ Ω m for 20-nm Cu, we obtain $R_{\rm Cu} = 6.2 \times 10^{-20}$ m$^2$. In order to quantitatively determine the intrinsic $g_{\rm YIG/Cu}^{\uparrow\downarrow}$, we grow a 2-μm (>> $\lambda_{\rm SD}$) Cu layer on YIG to reduce the backflow of spin current. Figure 3c shows the Gilbert damping enhancement of two YIG/Cu bilayers compared to a bare YIG film. The values of $\alpha_{\rm sp}$ for YIG/Cu(10 nm) and YIG/Cu(2 μm) are $(1.1 \pm 0.1) \times 10^{-4}$ and $(18 \pm 1) \times 10^{-4}$, respectively, clearly indicating significant backflow of spin current driven by spin accumulation when $t_{\rm Cu} << \lambda_{\rm SD}$. This is confirmed by FMR measurement shown in Fig. 3f, where the linewidth change for YIG/Cu(2 μm) is 8.7 Oe, much larger than the 1.2 Oe for YIG/Cu(10 nm). From the value of $\alpha_{\rm sp}$, we obtain $g_{\text{eff}}^{\uparrow\downarrow}$ = (3.4 ± 0.3) × $10^{18}$ $\text{m}^{-2}$ for the YIG/Cu(2 μm) bilayer. From Eqs. (2)-(4) and using $\epsilon$ = 1/68 and $\lambda_{\text{SD}}$ = 245 nm for Cu [31], we calculate the intrinsic $g_{\text{YIG/Cu}}^{\uparrow\downarrow}$ = (1.8 ± 0.2) × $10^{19}$ $\text{m}^{-2}$ for the YIG/Cu(2 μm) bilayer. As demonstrated in ref. 31, the $g_{\text{eff,trilayer}}^{\uparrow\downarrow}$ of the YIG/Cu(20 nm)/Pt and YIG/Cu(20 nm)/W trilayers can be obtained from the damping enhancement using Eq. (3) by replacing $g_{\text{eff}}^{\uparrow\downarrow}$ with $g_{\text{eff,trilayer}}^{\uparrow\downarrow}$, from which we calculate the spin conductance $g_{\text{Cu/Pt}}$ = (2.8 ± 0.4) × $10^{18}$ $\text{m}^{-2}$ which is smaller than $g_{\text{YIG/Pt}}^{\uparrow\downarrow}$ = (3.9 ± 0.3) × $10^{18}$ $\text{m}^{-2}$, and $g_{\text{Cu/W}}$ = (7.6 ± 0.9) × $10^{18}$ $\text{m}^{-2}$ which is much larger than $g_{\text{YIG/W}}^{\uparrow\downarrow}$ = (1.2 ± 0.1) × $10^{18}$ $\text{m}^{-2}$.

Using the spin mixing conductances and spin conductance calculated above for each interface in the YIG/Cu/Pt and YIG/Cu/W trilayers, we can understand the opposite behavior of spin currents shown in Figs. 2f and 2g. Figure 4 schematically shows the series circuits of spin current flow from YIG → Pt (or W) and from YIG → Cu → Pt (or W). Here, we use spin mixing resistance $R_{\text{YIG/NM}}^{\uparrow\downarrow} = 1/g_{\text{YIG/NM}}^{\uparrow\downarrow}$ to represent each YIG/NM interface and spin resistance $R_{\text{Cu/NM}}$ = $1/g_{\text{Cu/NM}}$ for each Cu/NM interface (Table I). For the YIG/Cu(20 nm)/Pt trilayer, the total spin resistance $R_{\text{YIG/Cu/Pt}}^{\uparrow\downarrow} = R_{\text{YIG/Cu}}^{\uparrow\downarrow} + R_{\text{Cu}} + R_{\text{Cu/Pt}}$ = (0.56 + 0.62 + 3.6) × $10^{-19}$ $\text{m}^2$ = 4.8 × $10^{-19}$ $\text{m}^2$, larger than $R_{\text{YIG/Pt}}^{\uparrow\downarrow}$ = 2.6 × $10^{-19}$ $\text{m}^2$ for YIG/Pt. As a result, the spin current at $t_{\text{Cu}} \geq 10$ nm is smaller (20-25%) than $J_s(0)$ for YIG/Pt (Fig. 2f). One notes that, relative to the YIG/Cu interface, the Cu/Pt interface is the dominant barrier to spin transport in the trilayer. For YIG/Cu(20 nm)/W, the total spin resistance $R_{\text{YIG/Cu/W}}^{\uparrow\downarrow} = R_{\text{YIG/Cu}}^{\uparrow\downarrow} + R_{\text{Cu}} + R_{\text{Cu/W}}$ = (0.56 + 0.62 + 1.3) × $10^{-19}$ $\text{m}^2$ = 2.5 × $10^{-19}$ $\text{m}^2$, smaller than $R_{\text{YIG/W}}^{\uparrow\downarrow}$ = 8.3 × $10^{-19}$ $\text{m}^2$ for YIG/W (Fig. 4b). This is why the spin current plateau for YIG/Cu/W is ~4.5 times larger than $J_s(0)$ for YIG/W as shown in Fig. 2g.

In summary, we systematically studied FMR spin pumping in YIG/Cu/Pt and YIG/Cu/W trilayers as well as YIG/Cu bilayers. Significant enhancement of spin currents is observed in YIG/Cu/W trilayers as compared to YIG/W bilayers with direct contact. From the spin pumping enhancement of Gilbert damping, we determined the spin mixing conductances of YIG/Pt, YIG/W, YIG/Cu interfaces, and spin conductances of Cu/Pt and Cu/W interfaces. These values explain the suppression of spin currents pumped into Pt in YIG/Cu/Pt trilayers and the enhancement of spin currents in YIG/Cu/W trilayers. This discovery potentially paves a path toward significant improvement of spin pumping efficiency by engineering multilayers with optimized spin conductance matching of the interfaces, a powerful capability for future spin-functional devices.

This work is supported by the Center for Emergent Materials at the Ohio State University, a NSF Materials Research Science and Engineering Center (DMR-0820414) (HLW and FYY) and by the Department of Energy through grant DE-FG02-03ER46054 (PCH). Partial support is provided by Lake Shore Cryogenics Inc. (CHD) and the NanoSystems Laboratory at the Ohio State University.

**Reference**

1. I. Žutić, J. Fabian, and S. Das Sarma, Spintronics: Fundamentals and applications, *Rev. Mod. Phys.* **76**, 323 (2004).
2. D. D. Awschalom and M. E. Flatté, Challenges for semiconductor spintronics, *Nature Physics* **3**, 153 (2007).
3. J. Chow, R. J. Kopp, and P. R. Portney, Energy Resources and Global Development, *Science* **302**, 1528 (2003).
4. Y. Tserkovnyak, A. Brataas, G. E.W. Bauer, and B. I. Halperin, Nonlocal magnetization dynamics in ferromagnetic heterostructures, *Rev. Mod. Phys.***77**, 1375 (2005).
5. Y. Tserkovnyak, A. Brataas, and G. E. W. Bauer, Spin pumping and magnetization dynamics in metallic multilayers, *Phys. Rev. B* **66**, 224403 (2002).
6. Y. Kajiwara, K. Harii, S. Takahashi, J. Ohe, K. Uchida, M. Mizuguchi, H. Umezawa, H. Kawai, K. Ando, K. Takanashi, S. Maekawa, and E. Saitoh, Transmission of electrical signals by spin-wave interconversion in a magnetic insulator, *Nature* **464**, 262 (2010).
7. B. Heinrich, C. Burrowes, E. Montoya, B. Kardasz, E. Girt, Y.-Y. Song, Y. Y. Sun, and M. Z. Wu, Spin pumping at the magnetic insulator (YIG)/normal metal (Au) interfaces, *Phys. Rev. Lett*. **107**, 066604 (2011).
8. K. Ando, S. Takahashi, J. Ieda, H. Kurebayashi, T. Trypiniotis, C. H. W. Barnes, S. Maekawa and E. Saitoh, Electrically tunable spin injector free from the impedance mismatch problem, *Nature Materials* **10**, 655 (2011).
9. M. V. Costache, M. Sladkov, S. M. Watts, C. H. van der Wal, and B. J. van Wees, Electrical detection of spin pumping due to the precessing magnetization of a single ferromagnet, *Phys. Rev. Lett.* **97**, 216603 (2006).

10. F. D. Czeschka, L. Dreher, M. S. Brandt, M. Weiler, M. Althammer, I.-M. Imort, G. Reiss, A. Thomas, W. Schoch, W. Limmer, H. Huebl, R. Gross, and S. T. B. Goennenwein, Scaling Behavior of the Spin Pumping Effect in Ferromagnet-Platinum Bilayers, *Phys. Rev. Lett*. **107**, 046601 (2011).
11. H. L. Wang, C. H. Du, Y. Pu, R. Adur, P. C. Hammel, and F. Y. Yang, Large spin pumping from epitaxial $Y_3Fe_5O_{12}$ thin films to Pt and W layers, *Phys. Rev. B* **88**, 100406(R) (2013).
12. C. H. Du, H. L. Wang, Y. Pu, T. L. Meyer, P. M. Woodward, F. Y. Yang, and P. C. Hammel, Probing the Spin Pumping Mechanism: Exchange Coupling with Exponential Decay in $Y_3Fe_5O_{12}$/barrier/Pt Heterostructures, *Phys. Rev. Lett.* **111**, 247202 (2013).
13. C. Hahn, G. de Loubens, O. Klein, M. Viret, V. V. Naletov, and J. Ben Youssef, Comparative measurements of inverse spin Hall effects and magnetoresistance in YIG/Pt and YIG/Ta, *Phys. Rev. B* **87**, 174417 (2013).
14. S. Mizukami, Y. Ando, and T. Miyazaki, Effect of spin diffusion on Gilbert damping for a very thin permalloy layer in Cu/permalloy/Cu/Pt films, *Phys. Rev. B* **66**,104413 (2002).
15. C. Burrowes, B. Heinrich, B. Kardasz, E. A. Montoya, E. Girt, Y. Sun, Y.-Y. Song, and M. Wu, Enhanced spin pumping at yttrium iron garnet/Au interfaces, *Appl. Phys. Lett.* **100**, 092403 (2012).
16. P. Deorani and H. Yang, Role of spin mixing conductance in spin pumping: Enhancement of spin pumping efficiency in Ta/Cu/Py structures, *Appl. Phys. Lett*. **103**, 232408 (2013).
17. K. Uchida, S. Takahashi, K. Harii, J. Ieda, W. Koshibae, K. Ando, S. Maekawa and E. Saitoh, Observation of the spin Seebeck effect, *Nature* **455**, 778 (2008).

18. T. Kikkawa, K. Uchida, Y. Shiomi, Z. Qiu, D. Hou, D. Tian, H. Nakayama, X.-F. Jin, and E. Saitoh, Longitudinal Spin Seebeck Effect Free from the Proximity Nernst Effect, *Phys. Rev. Lett.* **110**, 067207 (2013).

19. A. A. Serga, A. V. Chumak, B. Hillebrands, YIG magnonics, *J. Phys. D: Appl. Phys.* **43**, 264002 (2010).

20. Y. Sun, Y.-Y. Song, H. Chang, M. Kabatek, M. Jantz, W. Schneider, M. Wu, H. Schultheiss, and A. Hoffmann, Growth and ferromagnetic resonance properties of nanometer-thick yttrium iron garnet films, *Appl. Phys. Lett*. **101**, 152405 (2012).

21. S. M. Rezende, R. L. Rodríguez-Suárez, and A. Azevedo, Magnetic relaxation due to spin pumping in thick ferromagnetic films in contact with normal metals, *Phys. Rev. B* **88**, 014404 (2013).

22. S.Y. Huang, X. Fan, D. Qu, Y. P. Chen, W. G. Wang, J. Wu, T.Y. Chen, J. Q. Xiao, and C. L. Chien, Transport magnetic proximity effects in Platinum, *Phys. Rev. Lett.* **109**, 107204 (2012).

23. Y. Sun, H. Chang, M. Kabatek, Y.-Y. Song, Z. Wang, M. Jantz, W. Schneider, M. Wu, E. Montoya, B. Kardasz, B. Heinrich, S. G. E. te Velthuis, H. Schultheiss, and A. Hoffmann, Damping in Yttrium Iron Garnet Nanoscale Films Capped by Platinum, *Phys. Rev. Lett.* **111**, 106601 (2013).

24. H. Nakayama, M. Althammer, Y.-T. Chen, K. Uchida, Y. Kajiwara, D. Kikuchi, T. Ohtani, S. Geprägs, M. Opel, S. Takahashi, R. Gross, G. E. W. Bauer, S. T. B. Goennenwein, and E. Saitoh, Spin Hall Magnetoresistance Induced by a Nonequilibrium Proximity Effect, *Phys. Rev. Lett.* **110**, 206601 (2013).

25. H. L. Wang, C. H. Du, P. C. Hammel, and F. Y. Yang, Strain-Tunable Magnetocrystalline Anisotropy in Epitaxial $Y_3Fe_5O_{12}$ Thin Films, *Phys. Rev. B* **89**, 134404 (2014).

26. K. Ando, S. Takahashi, J. Ieda, Y. Kajiwara, H. Nakayama, T. Yoshino, K. Harii, Y. Fujikawa, M. Matsuo, S. Maekawa, and E. Saitoh, Inverse spin-Hall effect induced by spin pumping in metallic system, *J. Appl. Phys.* **109**, 103913 (2011).

27. E. Shikoh, K. Ando, K. Kubo, E. Saitoh, T. Shinjo, and M. Shiraishi, Spin-pump-induced spin transport in *p*-Type Si at room temperature, *Phys. Rev. Lett.* **110**, 127201 (2013).

28. S. S. Kalarickal, P. Krivosik, M. Wu, C. E. Patton, M. L. Schneider, P. Kabos, T. J. Silva, and J. P. Nibarger, Ferromagnetic resonance linewidth in metallic thin films: Comparison of measurement methods, *J. Appl. Phys.* **99**, 093909 (2006).

29. A. Kapelrud and A. Brataas, Spin Pumping and Enhanced Gilbert Damping in Thin Magnetic Insulator Films, *Phys. Rev. Lett.* **111**, 097602 (2013).

30. X. Jia, K. Liu, X. K, and G. E. W. B. Bauer, Spin transfer torque on magnetic insulators, *Europhys. Lett.* **96**, 17005 (2011).

31. B. Kardasz and B. Heinrich, Ferromagnetic resonance studies of accumulation and diffusion of spin momentum density in Fe/Ag/Fe/GaAs(001) and Ag/Fe/GaAs(001) structures, *Phys. Rev. B* **81**, 094409 (2010).

32. O. Mosendz, G. Woltersdorf, B. Kardasz, B. Heinrich, and C. H. Back, Magnetization dynamics in the presence of pure spin currents in magnetic single and double layers in spin ballistic and diffusive regimes, *Phys. Rev. B* **79**, 224412 (2009).

33. A. Ghosh, S. Auffret, U. Ebels, and W. E. Bailey, Penetration Depth of Transverse Spin Current in Ultrathin Ferromagnets, *Phys. Rev. Lett.* **109**, 127202 (2012).

**Table I**. Spin pumping parameters of bilayers and trilayers in this study: spin pumping enhanced damping $\alpha_{\mathrm{sp}} = \alpha_{\mathrm{YIG/NM}} - \alpha_{\mathrm{YIG}}$ or $\alpha_{\mathrm{sp}} = \alpha_{\mathrm{YIG/Cu/NM}} - \alpha_{\mathrm{YIG}}$, effective spin mixing conductance $g_{\mathrm{eff}}^{\uparrow\downarrow}$ or $g_{\mathrm{eff,trilayer}}^{\uparrow\downarrow}$, spin mixing conductance $g_{\mathrm{YIG/NM}}^{\uparrow\downarrow}$ at YIG/NM interfaces, spin conductance $g_{\mathrm{Cu/NM}}$ at Cu/NM interfaces, spin mixing resistance $R_{\mathrm{YIG/NM}}^{\uparrow\downarrow} = 1/g_{\mathrm{YIG/NM}}^{\uparrow\downarrow}$ at YIG/NM interfaces, spin resistance $R_{\mathrm{Cu/NM}} = 1/g_{\mathrm{Cu/NM}}$ at Cu/NM interfaces, and total spin mixing resistance $R_{\mathrm{trilayer}}^{\uparrow\downarrow} = 1/g_{\mathrm{eff,trilayer}}^{\uparrow\downarrow}$ for the trilayers.

| Structure | $\alpha_{\mathrm{sp}}$ | $g_{\mathrm{eff}}^{\uparrow\downarrow}$ ($\mathrm{m^{-2}}$) | $g_{\mathrm{eff,trilayer}}^{\uparrow\downarrow}$ ($\mathrm{m^{-2}}$) | $g_{\mathrm{YIG/NM}}^{\uparrow\downarrow}$ ($\mathrm{m^{-2}}$) | $g_{\mathrm{Cu/NM}}$ ($\mathrm{m^{-2}}$) | $R_{\mathrm{YIG/NM}}^{\uparrow\downarrow}$ or $R_{\mathrm{Cu/NM}}$ ($\mathrm{m^{2}}$) | $R_{\mathrm{trilayer}}^{\uparrow\downarrow}$ ($\mathrm{m^{2}}$) |
|---|---|---|---|---|---|---|---|
| YIG/Pt | $(2.1 \pm 0.1) \times 10^{-3}$ | $(3.9 \pm 0.3) \times 10^{18}$ | | $(3.9 \pm 0.3) \times 10^{18}$ | | $(2.6 \pm 0.2) \times 10^{-19}$ | |
| YIG/W | $(6.3 \pm 0.6) \times 10^{-4}$ | $(1.2 \pm 0.1) \times 10^{18}$ | | $(1.2 \pm 0.1) \times 10^{18}$ | | $(8.3 \pm 0.9) \times 10^{-19}$ | |
| YIG/Cu(2 μm) | $(1.8 \pm 0.1) \times 10^{-3}$ | $(3.4 \pm 0.3) \times 10^{18}$ | | $(1.8 \pm 0.2) \times 10^{19}$ | | $(5.6 \pm 0.5) \times 10^{-20}$ | |
| YIG/Cu(20 nm)/Pt | $(1.1 \pm 0.1) \times 10^{-3}$ | | $(2.1 \pm 0.3) \times 10^{18}$ | | | | $(4.8 \pm 0.7) \times 10^{-19}$ |
| YIG/Cu(20 nm)/W | $(2.1 \pm 0.2) \times 10^{-3}$ | | $(4.0 \pm 0.4) \times 10^{18}$ | | | | $(2.5 \pm 0.3) \times 10^{-19}$ |
| Cu/Pt | | | | | $(2.8 \pm 0.4) \times 10^{18}$ | $(3.6 \pm 0.5) \times 10^{-19}$ | |
| Cu/W | | | | | $(7.6 \pm 0.9) \times 10^{18}$ | $(1.3 \pm 0.2) \times 10^{-19}$ | |

**Figure Captions:**

**Figure 1**. (a) $\theta$-2$\theta$ XRD scan of a 20-nm thick YIG film on GGG (111), which exhibits clear Laue oscillations. Inset: rocking curve of the first satellite peak on the left of YIG (444) peak with FWHM of 0.0073°. AFM images of (b) a 20-nm bare YIG film (c) a YIG/Cu(5 nm) bilayer and (d) a YIG/Cu(20 nm) bilayers over an area of 10 μm × 10 μm with an rms roughness of 0.15, 0.19 and 0.22 nm, respectively.

**Figure 2.** (a) Schematic of experimental geometry for ISHE voltage measurements and $V_{ISHE}$ vs. $H$ - $H_{res}$ spectra of (b) a YIG/Pt bilayer, (c) a YIG/W bilayer, (d) a YIG/Cu(20 nm)/Pt trilayer and (e) a YIG/Cu(20 nm)/W trilayer at $\theta_H$ = 90° and 270° (in-plane fields). The thicknesses of all YIG, Pt and W layers are 20, 5 and 5 nm, respectively. Spin current $J_s$ in (f) YIG/Cu/Pt and (g) YIG/Cu/W trilayers normalized to $J_s(0)$ in YIG/Pt and YIG/W bilayers, respectively, as a function of the Cu thickness.

**Figure 3.** Frequency dependencies of FMR linewidths of (a) YIG/Cu/Pt, (b) YIG/Cu/W and (c) YIG/Cu series with corresponding FMR derivative absorption spectra at $f$ = 9.65 GHz shown in (d), (e) and (f).

**Figure 4.** Schematic comparison of spin mixing resistance or spin resistance of (a) YIG/Pt (red), YIG/Cu (brown) and Cu/Pt (green) interfaces, and (b) YIG/W (blue), YIG/Cu (brown) and Cu/W (purple) interfaces. The calculated values of spin mixing conductances explain the opposite behavior in Figs. 2f and 2g: the insertion of a Cu layer suppresses the magnitude of spin current from YIG to Pt, but enhances the spin current generation from YIG to W.

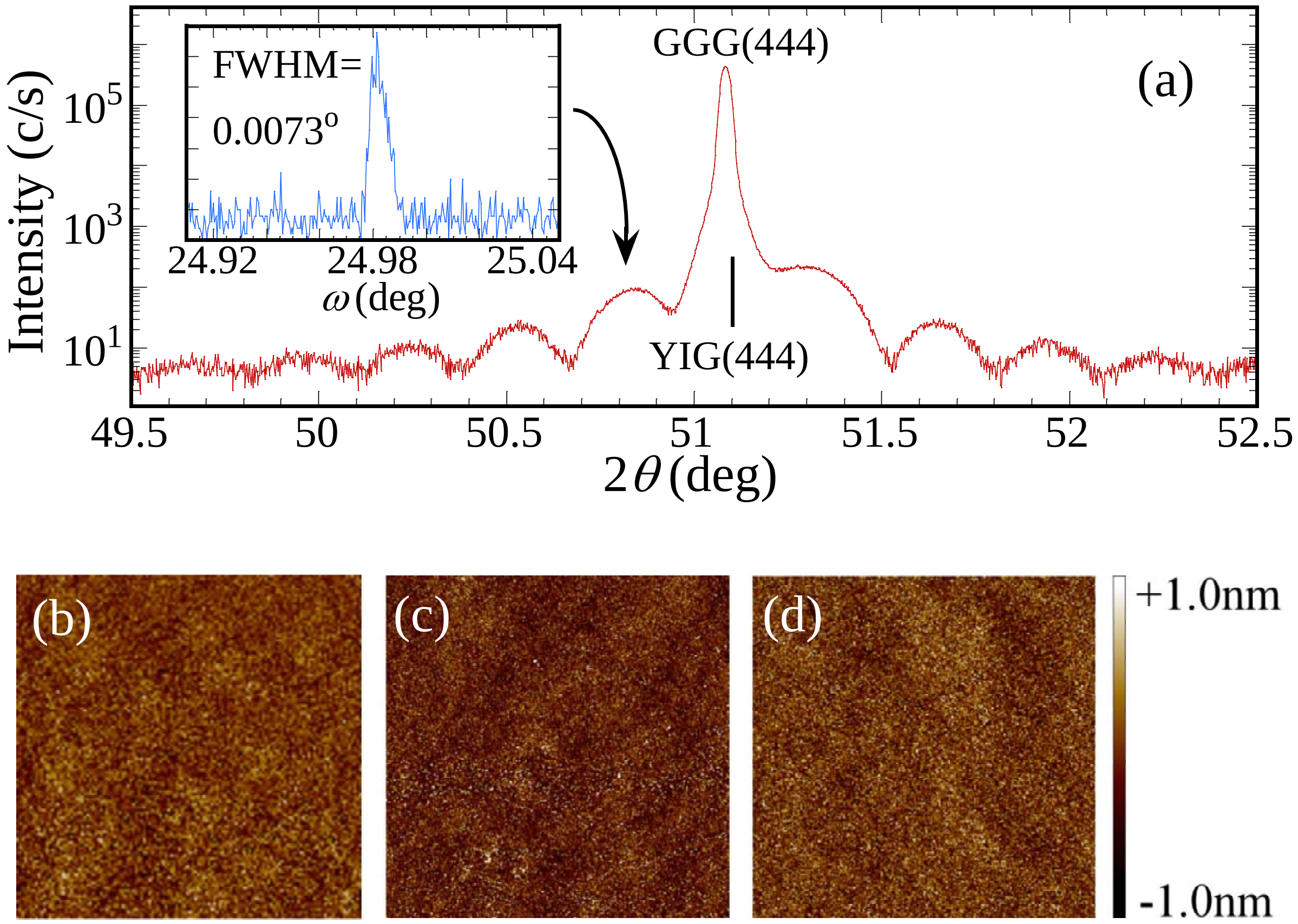

**Figure 1.**

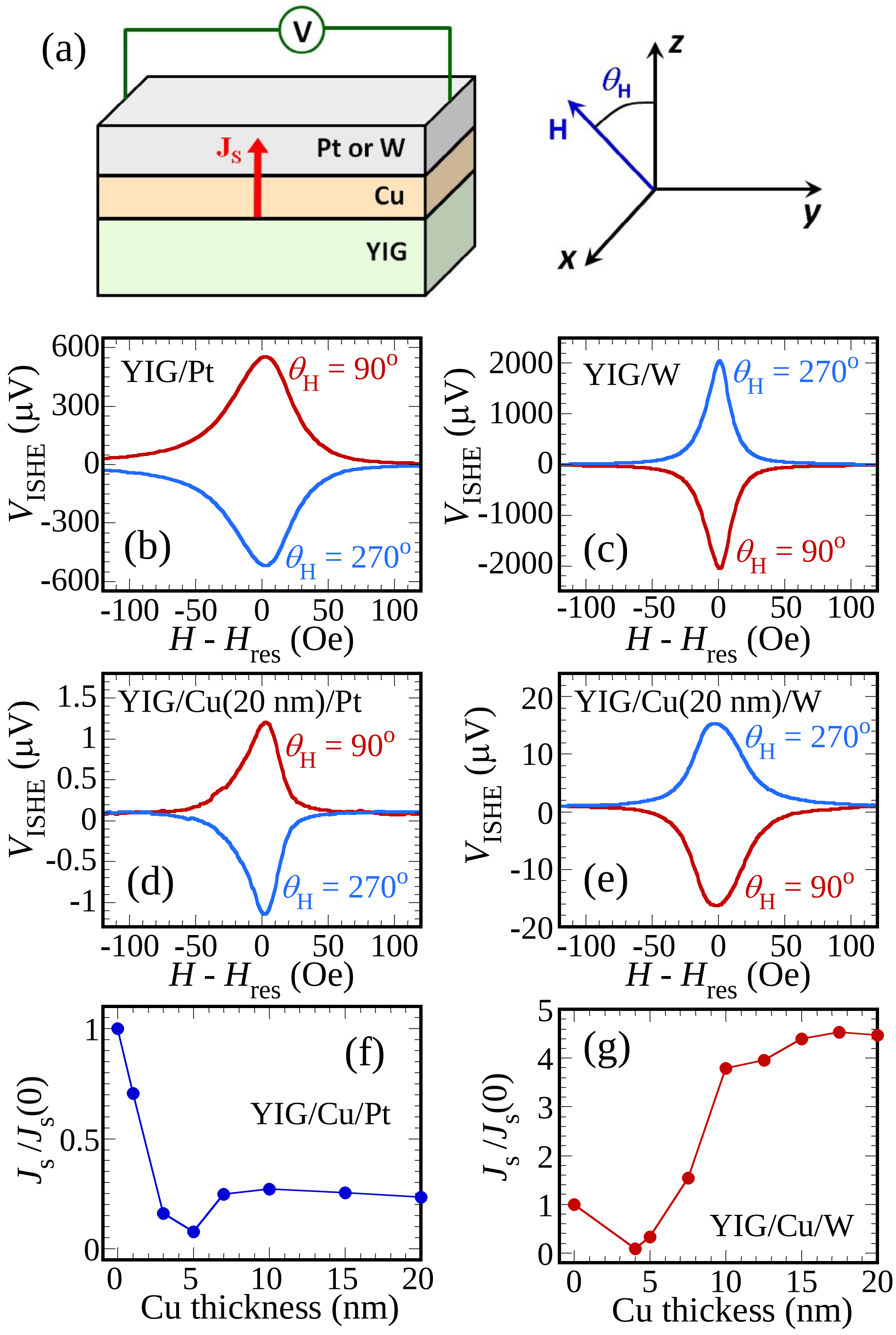

**Figure 2.**

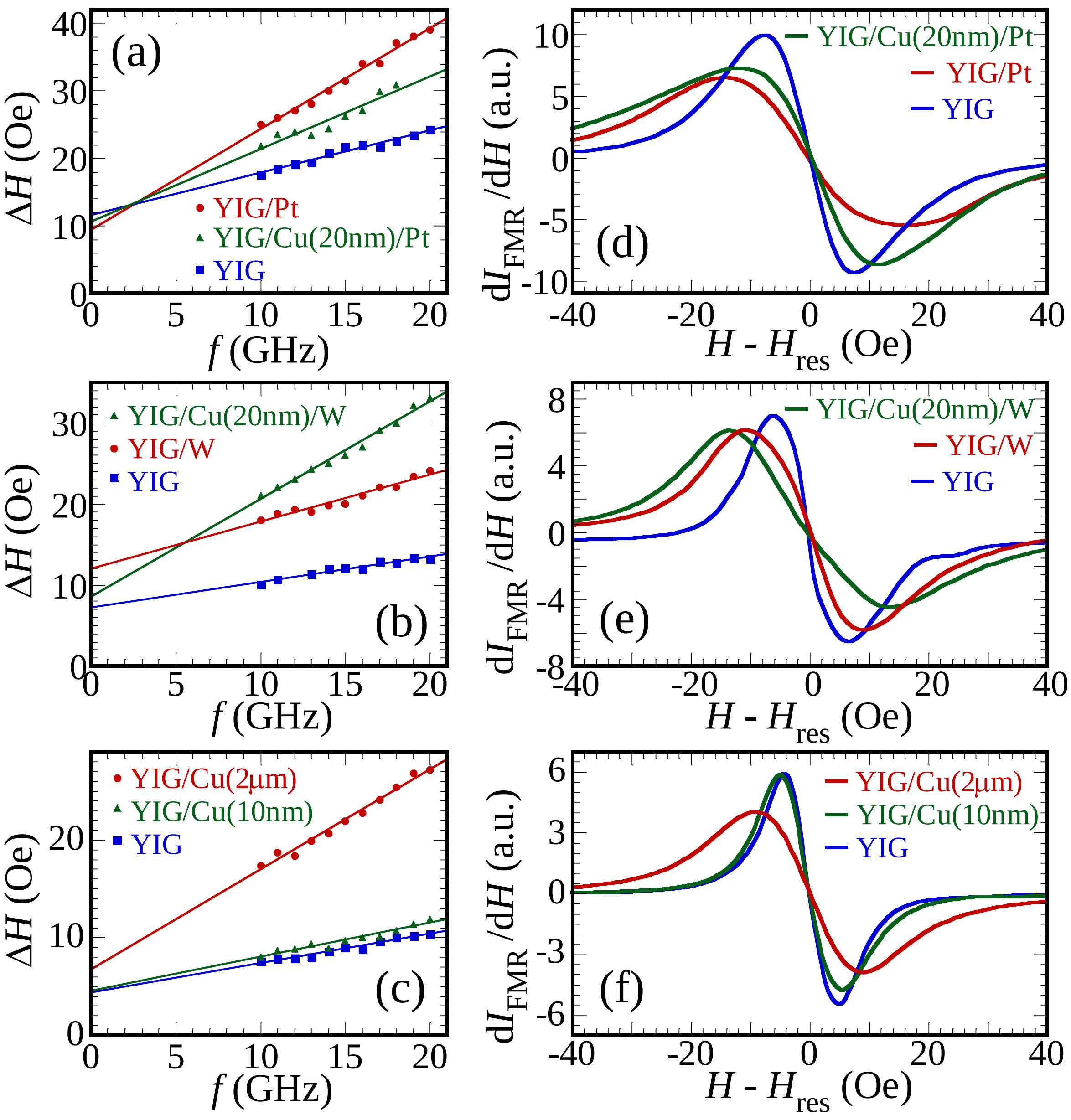

**Figure 3.**

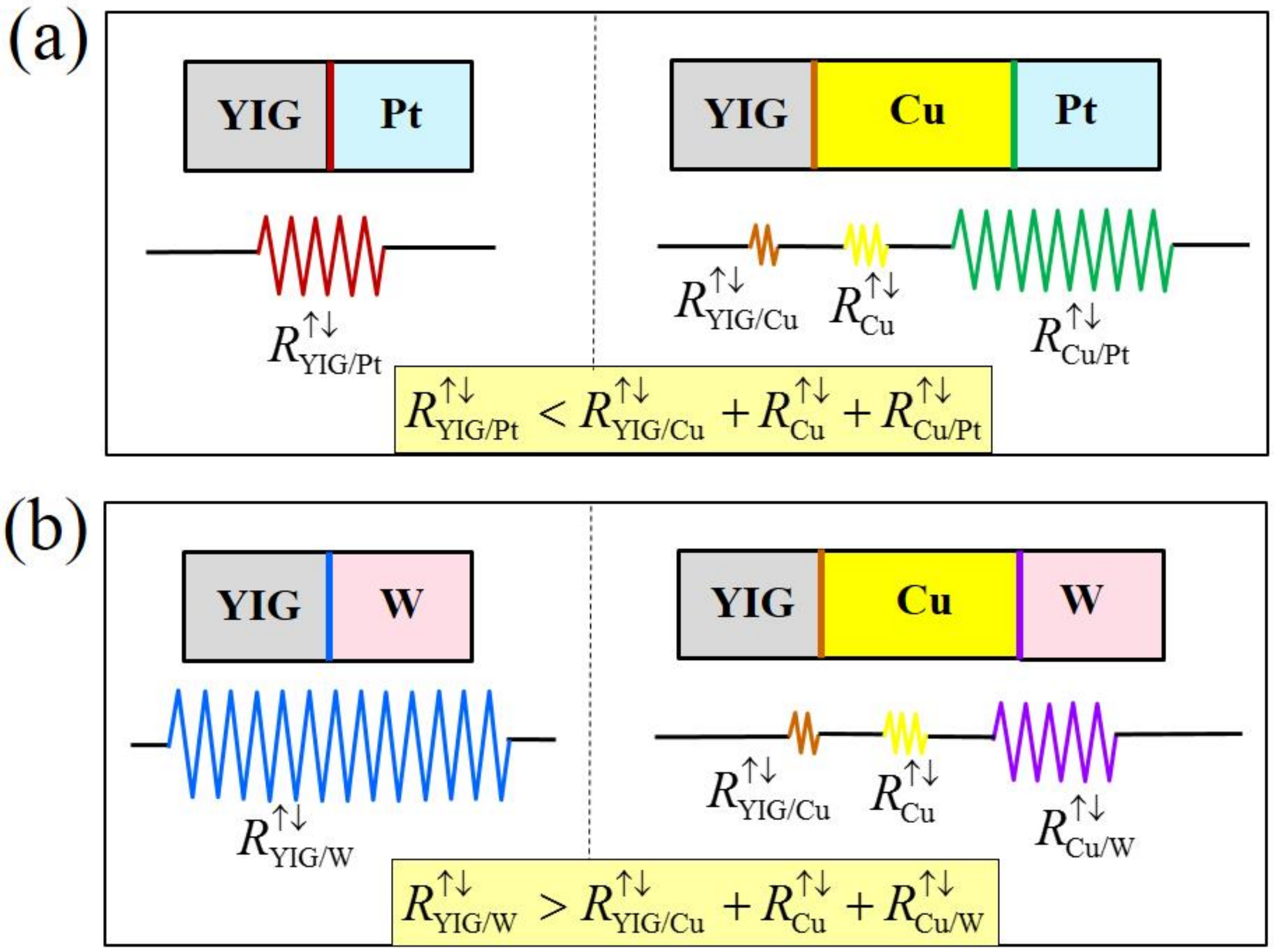

(a)
YIG
Pt
YIG
Cu
Pt
$R^{\uparrow\downarrow}_{\mathrm{YIG/Pt}}$
$R^{\uparrow\downarrow}_{\mathrm{YIG/Cu}}$
$R^{\uparrow\downarrow}_{\mathrm{Cu}}$
$R^{\uparrow\downarrow}_{\mathrm{Cu/Pt}}$
$R^{\uparrow\downarrow}_{\mathrm{YIG/Pt}} < R^{\uparrow\downarrow}_{\mathrm{YIG/Cu}} + R^{\uparrow\downarrow}_{\mathrm{Cu}} + R^{\uparrow\downarrow}_{\mathrm{Cu/Pt}}$
(b)
YIG
W
YIG
Cu
W
$R^{\uparrow\downarrow}_{\mathrm{YIG/W}}$
$R^{\uparrow\downarrow}_{\mathrm{YIG/Cu}}$
$R^{\uparrow\downarrow}_{\mathrm{Cu}}$
$R^{\uparrow\downarrow}_{\mathrm{Cu/W}}$
$R^{\uparrow\downarrow}_{\mathrm{YIG/W}} > R^{\uparrow\downarrow}_{\mathrm{YIG/Cu}} + R^{\uparrow\downarrow}_{\mathrm{Cu}} + R^{\uparrow\downarrow}_{\mathrm{Cu/W}}$

**Figure 4.**